\documentstyle[12pt,aaspp4]{article}

\def\pgamma{ \hat \gamma}
\def\gf2{(1+\epsilon)}
\def\gs{{\gamma^s}}
\def\bs{{\beta^s}}

\begin{document}

\title{Fluid dynamics of partially radiative blast waves}
\author{Ehud Cohen, Tsvi Piran and  Re'em Sari }
\affil{The Racah Institute of Physics, The Hebrew University, Jerusalem 91904, Israel }
\authoremail{udic@nikki.fiz.huji.ac.il}

\begin{abstract}
  We derive a self similar solution for the propagation of an extreme
  relativistic (or Newtonian) radiative spherical blast wave into a
  surrounding cold medium. The solution is obtained under the
  assumption that the radiation process is fast, it takes place
  only in the vicinity of the shock and that it radiates away a fixed
  fraction of the energy generated by the shock. In the Newtonian
  regime these solutions generalize the Sedov-Taylor adiabatic
  solution and the pressure-driven fully radiative solution. In the
  extreme relativistic case these solutions generalize the
  Blandford-McKee adiabatic solution. They provide a new fully
  radiative extreme relativistic solution which is different from the
  Blandford-McKee fully radiative relativistic solution.  This new
  solution develops a hot interior which causes it to cool faster than
  previous estimates. We find that the energy of the blast wave
  behaves as a power law of the location of the shock. The power law
  index depends on the fraction of the energy emitted by the shock.
  We obtain an analytic solution for the interior of the blast wave.
  These new solutions might be applicable to the study of GRB
  afterglow or SNRs.
\end{abstract}

\section{Introduction}
Many astrophysical phenomena (SNRs, $\gamma$-ray bursts - GRBs, AGN
hot spots, etc.)  are believed to involve  radiative shock
waves.  These shocks accelerate particles which emit the observed
radiation.  In particular, it is widely accepted that the recently
discovered GRBs afterglow results from an emission by relativistic
shocks, created by the interaction between an initial ejecta and the
interstellar medium. The recent observations of GRB
afterglow have lead to  numerous attempts to model this phenomena.

If the radiative mechanisms are slow compared to hydrodynamics time
scale, the blast wave evolution is adiabatic.  The propagation of 
such a  blast wave is described be a self similar solution:  The
Sedov-Taylor solution (\cite{sedov,taylor}) describes the Newtonian regime
and the Blandford \& McKee (1976) solution describes the extreme
relativistic regime.

We call the solution radiative if the radiative mechanisms are fast
compared to the hydrodynamic time scale.  A fully radiative solution
is one in which all the energy generated by the shock is radiated
away. \cite{OM88} have shown that if a fully
radiative blast wave emits its energy only in the vicinity of the
shock, it can be described by one of two possible self-similar
solutions:  The pressure driven snow-plow (PDS) or the momentum
conserving snow-plow (MCS) solution. In the PDS solution a thin shell
``snow plows'' through the external medium, driven by the pressure of
its hot, roughly isobaric, interior. In the MCS solution the interior
has been cooled and a thin shell slows down while conserving momentum.
\cite{bm} have found an extreme relativistic fully
radiative solution. This solution describes a thin shell with a cold
interior and it can be considered as the relativistic generalization
of the Newtonian MCS solution. However, momentum is not conserved in 
this solution, as in the relativistic case one has to consider the 
momentum carried by the emitted radiation. 

These solutions are either adiabatic or fully radiative. It is
likely that in some cases not all the energy would be radiated away,
even though the cooling is fast.  Such is the situation if the shock
distributes the internal energy between the electrons and protons and
there is no coupling between the two afterwards.  Since only the
electrons cool, only a fraction of the internal energy will be
radiated. It is likely that at least in the initial phases of a GRB
afterglow this would be the case. We consider here this ``partially
radiative'' scenario.

The goal of our paper is to find an analytic solution for the
evolution of Newtonian and relativistic partially radiative blast
waves. 
This, we believe, will
eventually lead to a physical description of the evolution of GRB
afterglows in their non-adiabatic stage, and of other astrophysical
phenomena. We find a self similar solution under the assumption that the
radiative mechanism is fast (compared to the hydrodynamics time
scale).  We parameterize the radiation by a dimensionless (the non
dimensionality is essential for a self-similar solution) parameter,
$\epsilon$, which describes the fraction of the energy produced by the
shock that is radiated away.  We recover the extreme relativistic
adiabatic Blandford-McKee solution and the Newtonian Sedov-Taylor
solution in the corresponding limit when $\epsilon = 0$.  In the
Newtonian case when $\epsilon \to 1$ we recover the PDS
solution, but we do not reproduce the MCS solution. Similarly in the
relativistic limit of $\epsilon \rightarrow 1$ the solution approaches
a new relativistic PDS solution which is different from the  radiative
Blandford-McKee solution.

We describe our model in Sec. \ref{s:model}. It is composed of an
adiabatic shock followed by a narrow radiative region and a wide
self-similar adiabatic regime.  We proceed in Sec.
\ref{s:newtonian_cond} by calculating the radiative shock conditions
for Newtonian blast waves. We describe the self similar equations and
their solution in Sec. \ref{s:newtonian_self}. Using the same method
we turn in Sec. \ref{s:rel} to the calculations of relativistic blast
waves.  Finally, in Sec. \ref{s:others_results} we discuss the
limiting solutions obtained by other authors and compare them with our
solution.

\section{The Model}
\label{s:model}
We consider a spherical radiative blast wave, that appears when a
large amount of energy is released within a small volume. This results
in a strong shock wave that expands supersonically into the
surrounding medium. We consider the regime where the influence of the
initial mass is negligible and that the pressure of the surrounding
medium is small compared to the energy density of the flow.  These
assumptions are necessary and sufficient to obtain a self similar
solution, see e.g.  \cite{ll}, pg. 404.  This solution corresponds in the
adiabatic limit to the well known Newtonian Sedov-Taylor solution or
to the relativistic Blandford-McKee solution.

We assume that the shocked, hot matter can radiate a fixed fraction of
its internal energy generated by the shock.  We show that the details
of the radiation mechanism are not important, as long as the cooling
time scale is short compared to the hydrodynamics time-scale.  We also
assume that the radiated energy is not re-absorbed in the system.
Under these assumptions we can divide the system into three parts. An
adiabatic shock, followed by a narrow radiative zone and a wide
adiabatic flow.

The assumption of fast cooling allows us to treat the shock and the
cooling layer as planar and stationary.  This means that the velocity
of the shock is constant during the time that a given fluid element
crosses the radiative zone and cools.  The physical conditions through
this ``radiative shock'' can be found using the equations of energy,
momentum and particle conservation.  For simplicity we assume that the
radiation does not alter the shock itself, which remains adiabatic,
and that a radiative layer follows it. However, the shock jump
conditions (Rankine-Hugoniot in the Newtonian case and Taub in the
relativistic case) are derived from the same conservation equations.
Thus, even though we divide the process into a shock and a radiative
layer we effectively use the conservation equations between the
unshocked matter and the shocked matter after it has been cooled.
This is valid even if the radiation process changes the shock itself
(i.e. in the case where the cooling length is comparable to the particles'
mean free path within the shock).

\begin{figure}
\begin{center}
\plotone{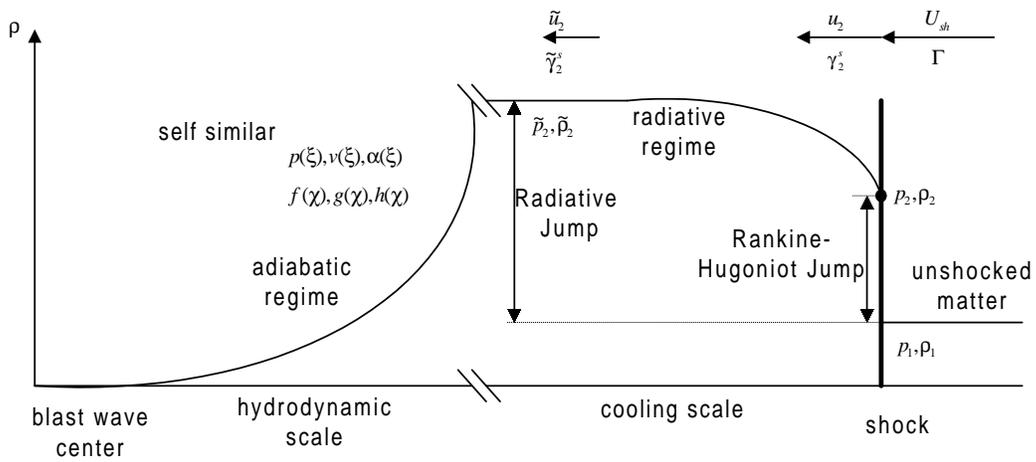}
    \caption{ Density as a function of distance from the center of the
      blast wave. The schematic drawing depicts the model used to
      obtain a self similar solution. It includes an adiabatic shock,
      a small radiative regime and an adiabatic self similar regime.}
    \label{schematic_draw}
  \end{center}
\end{figure}

Fig. \ref{schematic_draw} describes the model.  In the shock frame,
cold matter enters the shock. The shock itself is not affected by
radiation and can be described properly by the Newtonian
Rankine-Hugoniot or the relativistic Taub jump conditions.  The
downstream matter is hot and it moves sub-sonically relative to the
shock front. This matter cools by radiation and it is compressed by
the surrounding pressure. It stops radiating when it has emitted a
fixed fraction of its energy. It continues adiabatically afterwards.
The assumption of fast cooling enables us to treat the shock and the
radiation layer as instantaneous and with zero width compared to
the hydrodynamics scales. This translates to modified radiative jump
conditions relating the unshocked material and the cooled material
after the radiation zone. Using these conditions we find a self similar
solution for the remaining adiabatic interior.

\section{The Newtonian solution}

\subsection{Jump conditions of radiative shocks}
\label{s:newtonian_cond}
We divide the radiative shock, as defined in Sec. \ref{s:model}, to two
regions: The  ``adiabatic'' shock, and the 
radiative zone.  The conditions for the matter
immediately after the shock are given by the Rankine-Hugoniot jump
conditions. Assuming that the matter is adequately described by a
polytropic equation of state with an adiabatic index $\pgamma$, the
jump conditions in the shock frame are (see e.g. \cite{ll} pg. 335) :
\begin{equation}
\label{n_shock}
\rho_2 = \left( { \pgamma + 1 \over \pgamma - 1 } \right) \rho_1,
\quad 
u_2 = \left( { \pgamma - 1 \over \pgamma + 1 } \right) U_{sh}, 
\quad
p_2 = { 2 \over \pgamma+1 } \rho_1 U_{sh}^2,
\end{equation}
where $\rho_1$ is the density of the unshocked matter, $U_{sh}$ is the
velocity of the shock relative to the unshocked matter and $p_2$,
$\rho_2$ and $u_2$ are the pressure, density and velocity of the
shocked matter, measured in the shock frame.  These jump conditions
hold if the shock is strong, i.e. the pressure of the unshocked gas is
negligible $(p_1 \ll \rho_1 U_{sh}^2)$.

In the radiative zone, the matter that was heated by the shock
radiates and  cools. In the mean time  it is also compressed by the surrounding pressure.
This compression causes the gas to slow down in order to conserve the
matter flux. At some point the matter reaches some equilibrium and
stops radiating.

To calculate the hydrodynamics of the blast wave interior, we need to
know the pressure, velocity and density of matter at the end of the
radiative layer.  Since we assume the cooling is fast, the shock
velocity is constant during the cooling of a fluid element of shocked
matter. We look therefore for a steady solution for the conditions of
the matter during the cooling process. We find the conditions at the
end of the radiative layer as a function of energy lost through this
process.  

In a steady state we can use, in the shock frame, the
equations of conservation of matter:
\begin{equation}
\label{n_1d_part}
\rho u = {\rm constant},
\end{equation}  
and of  momentum:
\begin{equation}
\label{n_1d_mom}
(\rho u) u + p = {\rm constant},
\end{equation}
where $\rho, u$ and $p$ are the fluid's density, velocity and
pressure.  These two conservation equations contain three unknowns, so
that we can find the physical quantities as a function of one another.
We solve equations \ref{n_1d_part} and \ref{n_1d_mom} for $\rho$ and
$p$ as function of $u$,
\begin{equation}
\rho(u) = { \rho_2 u_2 \over u }, \quad p(u) =  \rho_2 u_2  ( u_2 - u ) + p_2,
\end{equation}
where $\rho_2,u_2$ and $p_2$, the density, the velocity and the
pressure of the flow immediately after the shock, are given by Eq.
\ref{n_shock}.

\begin{figure}
\begin{center}
\plotone{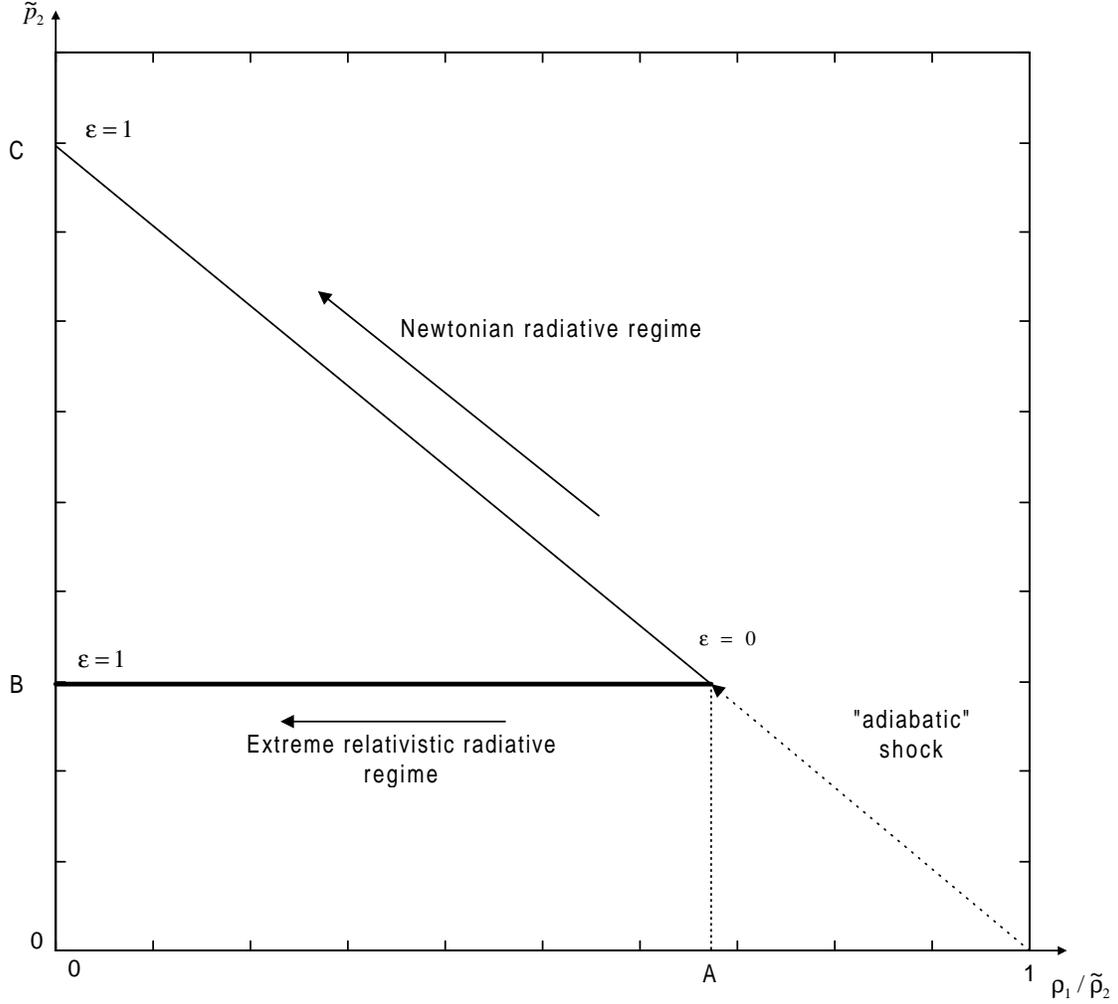}
\caption{ 
  Pressure as a function of density for the shock and the cooling
  profile.  The dotted arrow illustrates the adiabatic shock. The
  fluid then follows the appropriate line ( Newtonian - thin; Extreme
  relativistic - thick) until it reaches the cooling parameter.
  Labels A and B describes the density and pressure of the shocked matter immediately after
  the shock ($\rho_2$, $p_2$). Note that in the relativistic case the
  pressure is constant within the cooling layer. Label C describes the
  maximal pressure ($\rho_1 U_{sh}^2$), reached in the fully radiative
  Newtonian case at the end of
  the cooling layer.
\label{P_vs_V}
}
\end{center}
\end{figure}

We denote by $\tilde \rho$, $\tilde u$ and $\tilde p$ the density,
the velocity and the pressure when the fluid has stopped radiating, and
use the shock conditions ( Eq. \ref{n_shock} )  as the
boundary conditions for the radiative flow. 
We use the velocity at the end of the cooling layer to parameterize 
the fraction of energy lost, and find the pressure at that
location by
\begin{equation}
\label{n_tld_p} 
\tilde p_2 = p_2 ( 1 + { \pgamma-1 \over 2} \delta ) = { 2 + \delta(
  \pgamma-1) \over \pgamma+1} \rho_1 U_{sh}^2,
\end{equation}
the velocity relative to the observer, 
\begin{equation}
\label{n_tld_u}
U_{sh} - \tilde u_2 = (U_{sh} - u_2)( 1 + { \pgamma-1 \over 2} \delta
) =  { 2 + \delta(
  \pgamma-1) \over \pgamma+1} U_{sh},
\end{equation}
and the density is
\begin{equation}
\label{n_tld_r}
\tilde \rho_2 = { \rho_2  \over  1 -  \delta }= { \pgamma+1 \over
  (\pgamma-1) (1-\delta) } \rho_1,
\end{equation}
where $\delta \equiv 1 - \tilde u_2 / u_2$. Fig. \ref{P_vs_V} shows
the trajectory of a fluid element during the shock and the following
cooling process.


During the cooling process energy is radiated away. We write the energy
flux as                                                                          
\begin{equation}
\label{n_1d_ener}
{\cal J} =   u ( \rho u^2 /2 + e + p ),
\end{equation}
where the energy per unit volume is denoted by $e$. We find that the
fraction of energy flux lost in the radiative layer is:
\begin{equation}
\label{n_epsilon}
\epsilon(\delta) = 1 - {  {\cal J}(\tilde u_2) \over {\cal J}(u_2) } = 
{ \delta \over 1+\pgamma }\left[{2+ (\pgamma-1)\delta}\right].
\end{equation}

Note that we have not used the details of the cooling process to find
the pressure and density as a function of the velocity. This
is sufficient for our calculations, because all we need
are the physical conditions at a location were the matter has stopped
cooling. The velocity at this point is given by Eq.
\ref{n_epsilon}. 
\footnote{
The cooling profile is determined by the radiation mechanism. It is
only the conditions at the end of the cooling process that are
independent of this mechanism.}

Using Eqs. \ref{n_1d_ener} and \ref{n_epsilon} we 
calculate the overall energy loss rate: 
\begin{equation}
  \label{n_dE_dt}
  {dE \over dt} = - 4 \pi R_{sh}^2 {{\cal J}_{sh}} \epsilon = 
- 2 \pi \rho_1 U_{sh}^3 R_{sh}^2 \epsilon,
\end{equation}
where ${{\cal J}_{sh}}= \rho_1 U_{sh}^3 / 2$ is the energy flux of 
the matter entering the shock in shock frame.

\subsection{The self similar solution}
\label{s:newtonian_self}

After the matter has passed the radiation layer, it continues to flow
adiabatically.  The motion is described by matter, momentum and energy
conservations equations in spherical coordinates:
\begin{equation}
\label{s_part}
{ \partial \rho \over \partial t} + { 1 \over r^2 }  { \partial  \over
  \partial r} ( r^2 \rho u ) = 0, 
\end{equation}
\begin{equation}
\label{s_mom}
{ \partial u \over \partial t} + u  { \partial u  \over \partial r}  =
- { 1 \over \rho} { \partial p \over \partial r}, 
\end{equation}
and
\begin{equation}
\label{s_ener}
{ \partial  \over \partial t} [ e + \rho u^2/2 ] 
 + { 1 \over r^2 }  { \partial  \over \partial r}[ r^2  u ( e+p +\rho
 u^2/2)]  = 0.
\end{equation}

We look for a self similar solution for these equations.
Similarly to
the non-radiative Sedov-Taylor solution we consider the case where
$\rho_1 U_{sh}^2$, the momentum flux of the matter that enters the
shock is much larger than the thermal pressure $p_1$ of the
undisturbed medium.  

In the Sedov-Taylor solution dimensional
arguments lead to a self-similar dimensionless variable. In the
radiative case the definition of the self similar variable is not
straight forward, because the energy is not constant in time.
We look for a
self similar solution to the hydrodynamics equations with energy that  
varies with time as a power-law:
$E=E_0 ( t / t_0 )^{\lambda}$. We define a dimensionless variable
\begin{equation}
\label{n_ss}
\xi = r \left[ { \rho_1 \over t^{2+\lambda} (E_0 / t_0^{\lambda})} \right ]^{1/5}.
\end{equation} 
If this variable leads to a self similar solution, $r$ and
$t$  will appear in the solution only through $\xi$, and the
position
of the shock  corresponds to a fixed value of $\xi$, denoted 
$\xi_0$. Thus:
\begin{equation}
  \label{n_shell}
  R_{sh}(t)=\xi_0 \left [ {  t^{2+\lambda} (E_0 / t_0^{\lambda}) \over \rho_1}
\right ]^{1/5},
\quad
U_{sh}(t) = {d R_{sh}(t) \over dt}=  {2+\lambda \over 5} { R_{sh} \over t}.
\end{equation}
We substitute $R_{sh}$ and $U_{sh}$ into Eq. \ref{n_dE_dt} and obtain 
a cubic equation for $\lambda$ (which has only one real solution):
\begin{equation}
\label{n_beta_xi}
\lambda = - 2 \pi ({ 2+\lambda \over 5 })^3 \xi_0^5 \epsilon.
\end{equation}

Using the dimensionless similarity variable $\xi$, we look for a self
similar solution of
Eqs. \ref{s_part},\ref{s_mom},\ref{s_ener} of the 
form:
\begin{equation}
\rho(r,t) =  \rho_2 \alpha(\xi),
\end{equation}
\begin{equation}
u(r,t) = (U_{sh} - u_2) {r \over R_{sh}(t)} v(\xi),
\end{equation}
\begin{equation}
p(r,t) = p_2 ({r \over R_{sh}(t)})^2 p(\xi).
\end{equation}
The coefficients in these equations are chosen in order to match the
standard definitions in the adiabatic case (see e.g. \cite{shu}). 
 Using Eqs. \ref{n_tld_p}, \ref{n_tld_u}, \ref{n_tld_r} we
obtain the boundary conditions:
\begin{equation}
\label{n_ss_initial}
\alpha(\xi_0)={1 \over  1 -  \delta}, \quad v(\xi_0) =  1 + {
  \pgamma-1 \over 2} \delta ,
\quad p(\xi_0) = 1 + { \pgamma-1 \over 2} \delta.
\end{equation}

To determine the dimensionless position of the shock front, $\xi_0$,
we require that  
the energy in the blast wave interior equals with the energy defined by the self similar 
variable:
 \begin{equation}
\label{n_E_of_t}
E(t) = E_0 ({ t \over t_0})^{\lambda} = \int_0^{R_{sh}(t)} ( { p(r) \over \pgamma -1 } + { \rho(r) u(r)^2 \over  2}) 4 \pi r^2 dr. 
\end{equation}
Substitution of the self-similar functions into this equation yields
a non-dimensional normalization equation:
\begin{equation}
( {2+\lambda \over 5})^2 { 8 \pi \over \pgamma ^2-1}  \int_0^{\xi_0} (  p(\xi)  + { \alpha(\xi) v(\xi)^2 } )  \xi^4 d\xi = 1. 
\end{equation}

\begin{figure}
\begin{center}
\plotone{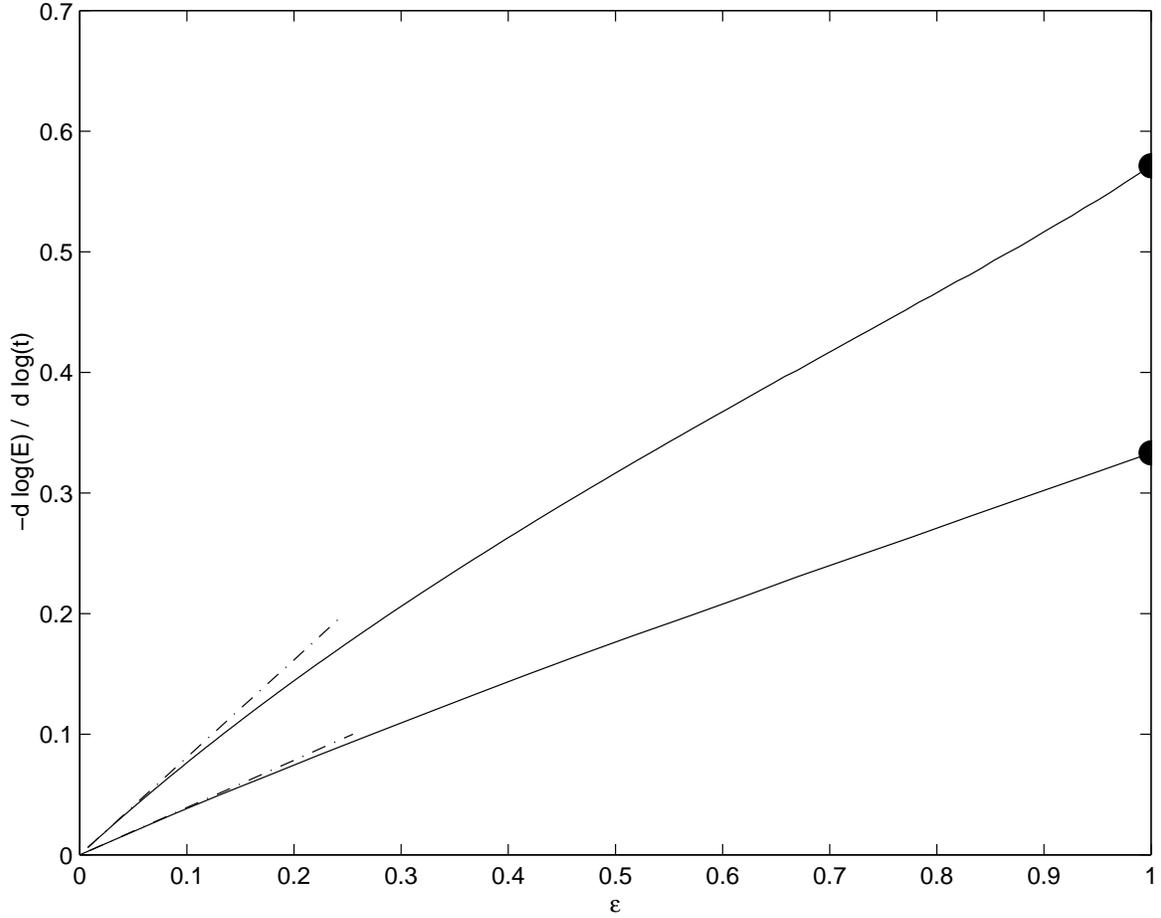}
\caption{ 
The energy loss rate  $- d \log(E) / d \log(t)$  as a function of $\epsilon$ for $\pgamma=4/3$
(lower curve)  and
$\pgamma=5/3$ (upper curve).
The dashed-dotted lines are the linear approximation for $\epsilon \to 0$,
assuming that the self similar profile does not depend on
$\epsilon$. The  circles are the corresponding PDS solutions.
\label{loss_newt}
}
\end{center}
\end{figure}

\begin{figure}
\begin{center}
\plotone{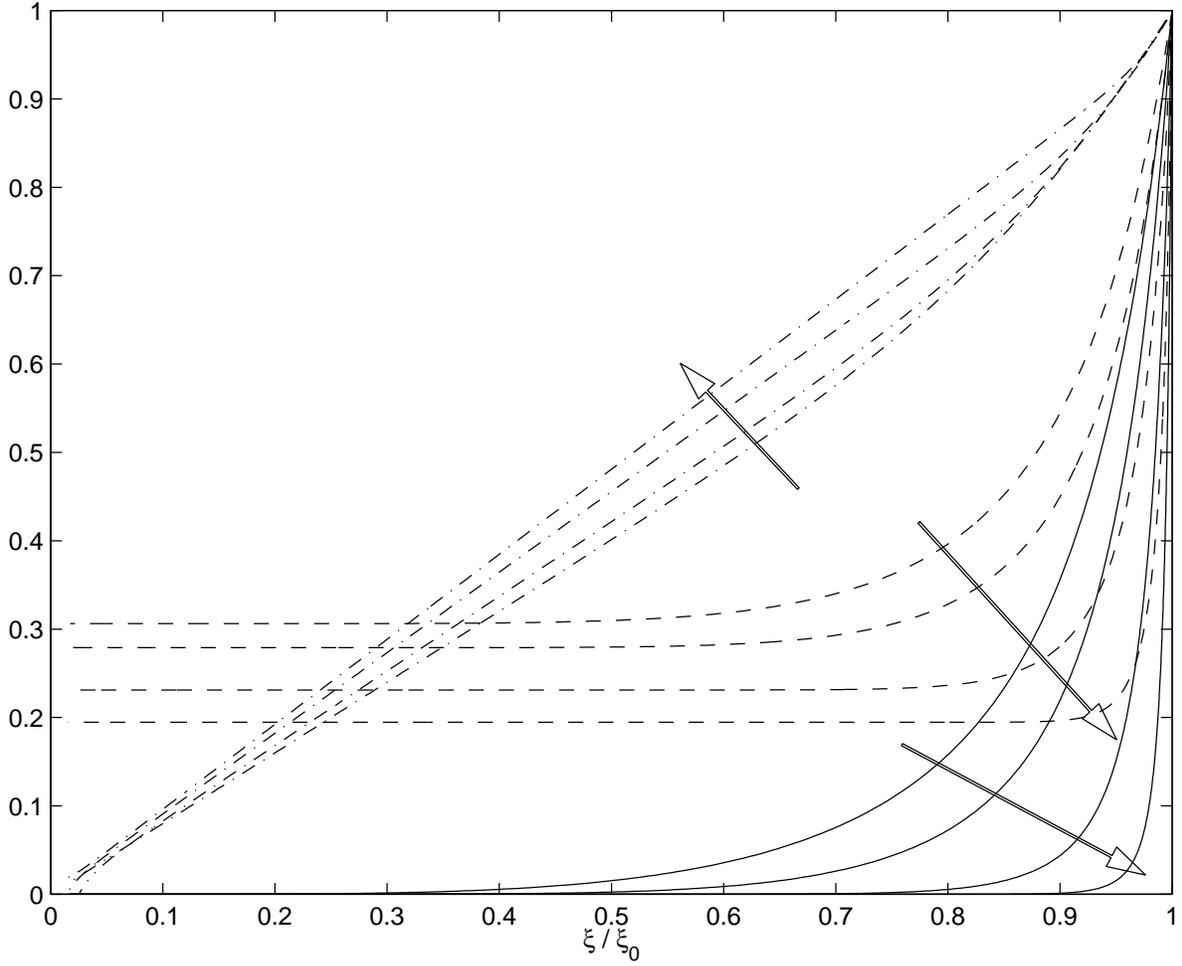}
\caption{ 
The normalized density (solid), velocity(dashed-dotted)  and pressure
(dashed)  profiles of the self similar Newtonian solution, for
$\epsilon=0,0.3,0.7,0.9$. The arrows show the direction of increasing $\epsilon$.
\label{newt_profile}
}
\end{center}
\end{figure}

Finally we substitute the self-similar functions $\alpha(\xi)$, $v(\xi)$ and $p(\xi)$ into the
fluid equations \ref{s_part}-\ref{s_ener}, replacing the partial derivatives 
with the corresponding total derivatives:
\begin{equation}
{ \partial \over \partial r} = { \xi \over r } { d \over d \xi}, \quad { \partial \over \partial t} = - { 2+\lambda \over 5 } { \xi \over t } { d \over d \xi}.
\end{equation}
We obtain the coupled ordinary differential equations:
\begin{eqnarray}
{{\cal D} \over 2} { d \log \alpha \over d \log \xi} &=& \Big( \alpha v (1+\pgamma-2 v)   
  \left[ (1+3 \lambda) (1+\pgamma)-4 (2+\lambda) v \right] + \\ \nonumber
& & 2 (\lambda-3) (\pgamma^2-1)p \Big) / (1+\pgamma-2 v) , \\
 {\cal D}{ d \log v \over d \log \xi} &=&  -\alpha (1+\pgamma-2 v) \left[ 5
    (1+\pgamma)-2 (2+\lambda) v \right]+ \\ \nonumber
& & 2 p (\pgamma-1) \left[{ (\lambda-3) (1+\pgamma)+3 (2+\lambda) \pgamma v} \right],  \\
{{\cal D} \over 2} { d log p \over d \log \xi} &=&   2 (2+\lambda)
(\pgamma-1) \pgamma p+\alpha(1+\pgamma) \times    
 \\ \nonumber
& & \left[ (14+2 \lambda+\pgamma+3 \lambda \pgamma) v -5 (1+\pgamma)-4 (2+\lambda) {{v}^2} \right] ,
\end{eqnarray}
where
\begin{equation}
  {\cal D}= \left( {2 (1-\pgamma) \pgamma p+\alpha {(1+\pgamma-2
        v)}^2}\right) ( 2 + \lambda).
\end{equation}
We solve these equations numerically, using the boundary conditions
from Eq. \ref{n_ss_initial}, and  obtain the relation between the cooling
parameter $\epsilon$ and the power-law index $\lambda$ (see
Fig. \ref{loss_newt}). Self similar solutions for several cooling
parameters are shown in Fig. \ref{newt_profile}.  

For  $\epsilon=0$ our solution is adiabatic and we recover the
classical Sedov-Taylor solution.
For $\epsilon=1$ the density diverges (Eq. \ref{n_tld_r}) and the
boundary conditions become singular. Thus, to obtain the fully
radiative limit we must take the limit of  $\epsilon \to 1$.
We discuss this limit of the Newtonian and the relativistic solutions 
in section \ref{s:eps_1}.

For intermediate values of $\epsilon$ we see (Fig. \ref{newt_profile}) 
that as the cooling parameter becomes
larger, the  matter  concentrates in a small shell near the shock and the
internal pressure decreases. 
Notice that similarly to the non-radiative Sedov-Taylor solution, 
the interior temperature $T \propto P/
\rho$ diverges at the center of the blast
wave and it monotonously decreases towards the shock. This behavior
occurs because the material near the center passed through the front
earlier, when the shock speed was larger ( infinitely large in the
formal extrapolation to $t \to 0$), and it cools slowly, adiabatically.

\section{Extreme relativistic solution}
\label{s:rel}
The model described in Sec. \ref{s:model} is adequate for relativistic
blast waves as well as for Newtonian ones.  We assume 
that each shocked particle
emits a fraction of the internal energy it acquired in the shock.  We
also assume that the radiation time scale is short compared to the
hydrodynamic time scale, thus allowing us to treat the blast wave as
stationary when calculating the radiation process.

We start by calculating the conditions at the end of the ``radiative
shock'' as a function of the fraction of energy emitted. Then we use
these results as boundary conditions for the blast wave, assuming that
the flow outside of the radiation layer is adiabatic.
Following \cite{bm} we derive a self similar solution for extreme
relativistic blast wave, accurate to the leading order in
$\Gamma^{-2}$.
However, within the derivation the second order terms in the equations are also
important. 

\subsection{Jump conditions of radiative shocks}

We begin by calculating the jump conditions for a radiative extreme
relativistic strong shock ($\Gamma \gg 1$) using the relativistic
equation of state $\pgamma=4/3$.
The jump conditions are derived  from the
continuity of the energy, momentum and particle flux densities in the
shock frame. (see e.g. \cite{ll} pg. 511 ). Assuming that the unshocked matter
is cold, $p_1 \ll \rho_1 \Gamma^2$, we have 
\begin{equation}
e_2 = 2 \Gamma^2 \rho_1,  
\end{equation}
\begin{equation}
\rho_2 = 2 \sqrt{2} \rho_1 \Gamma, 
\end{equation}
\begin{equation}
\gamma_2^2 = { 1 \over 2}  \Gamma^2.
\end{equation}


For generality we allow for any polytropic index, and relax our
assumption of extreme relativistic flow when calculating the cooling
profile. Later on, while tying together the shock and the cooling
layer, we return to an extreme relativistic motion. 

In a relativistic flow the thermal energy can be comparable
to the rest mass energy, and we have to take into account the momentum
of the radiation. We cannot calculate the profile from matter and
momentum flux conservation alone as in the Newtonian case, and the
independence of the cooling profile from the nature of the radiation mechanism
is not obvious.  We assume that the radiation is emitted isotropically
in the fluid rest frame with a local cooling function ${\cal
  L}(e,\rho)$ (like in the Newtonian case this function does not
appear in the final result), and
calculate the cooling profile in the shock frame using matter
conservation:
\begin{equation}
\label{r_1d_particle}
{ \partial \over \partial x} \gs \bs \rho = 0,
\end{equation}
momentum conservation:
\begin{equation}
\label{r_1d_momentum}
{ \partial \over \partial x} [  \gs^2 \bs^2  ( \rho + \pgamma e )
+ ( \pgamma - 1 ) e ] = - 2 { {\cal L}(e,\rho) \over \bs} \gs^2 \bs
\end{equation}
and 
energy conservation:
\begin{equation}
\label{r_1d_energy}
{ \partial \over \partial x}  \left[{ \gs^2 \bs ( \rho + \pgamma e)}\right]   =
-{ {\cal L}(e,\rho) \over \bs } \gs^2 (1+\bs^2).
\end{equation}

To solve these equations we rewrite  Eq. \ref{r_1d_particle} as
\begin{equation}
\label{r_tilde_rho2}
\rho = { \rho_2 \gamma_2^s \beta_2^s \over \gs \bs }.
\end{equation}
The cooling function ${\cal L}$ can be eliminated from 
Eqs. \ref{r_1d_momentum} and \ref{r_1d_energy}, which after a little algebra and 
the usage of Eq. \ref{r_tilde_rho2} results in:
\begin{equation}
{ d e \over     d \bs} = - { \beta_2^s \gamma_2^s \rho_2 \gs \over  \pgamma -1  
-{\bs}^2}.
\end{equation}
Integrating this differential equation we obtain 
\begin{equation}
\label{r_tilde_e2}
e = e_2 + { \beta^s_2 \gamma_2^s \rho_2 \over 2 \sqrt{ (2-\pgamma)(
    \pgamma -1)}  } 
\log \left [ \Phi(\bs) / \Phi(\beta^s_2)
\right],  
\end{equation}
where
\begin{equation}
\Phi(\beta) = { (1 - \beta / \sqrt{\pgamma-1} ) ( 1 + 
\beta \sqrt{\pgamma-1}  + \sqrt{ 2-\pgamma} / \gamma )  \over 
(1 + \beta / \sqrt{\pgamma-1} ) ( 1 - 
\beta \sqrt{\pgamma-1}  + \sqrt{ 2-\pgamma} / \gamma ) 
} \nonumber.
\end{equation}

\begin{figure}
\begin{center}
\plotone{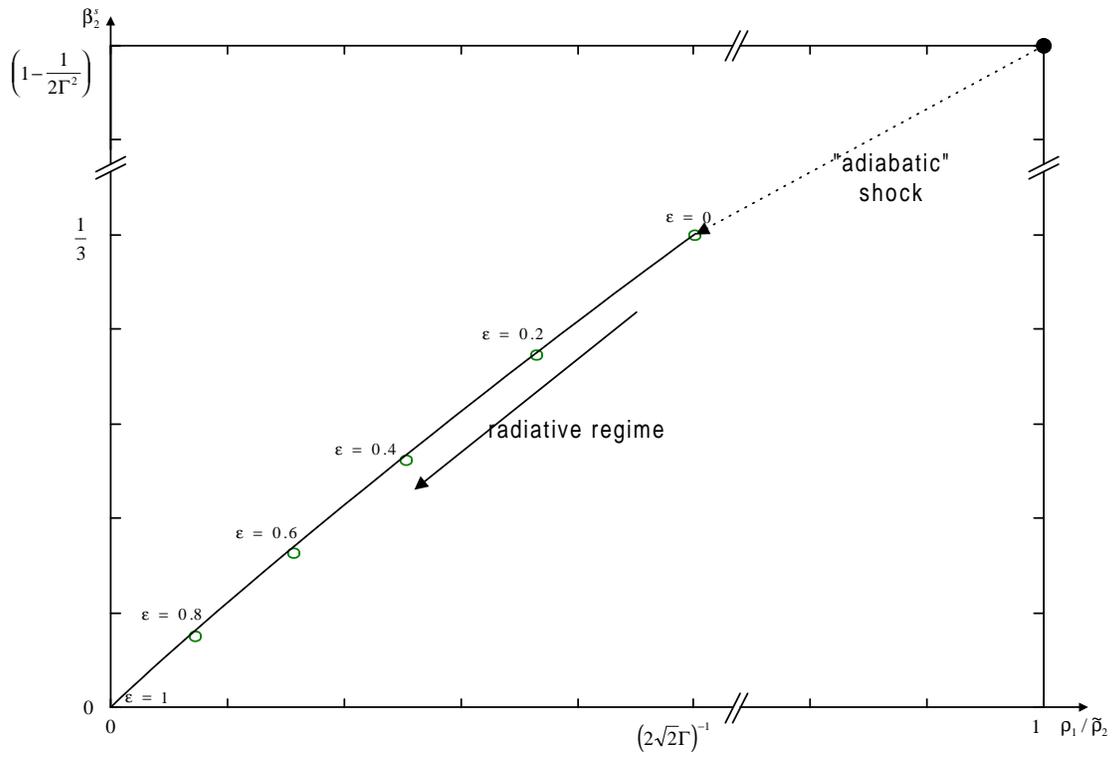}
\caption{ 
The velocity in the shock frame as a function of the density for
the extreme relativistic shock (dotted arrow) and for the cooling profile (solid line)
\label{beta_vs_V}
}
\end{center}
\end{figure}

Fig. \ref{beta_vs_V} depicts the flow of matter
through the ``radiative shock''. In the shock frame the dense shocked matter 
flows out of the shock with velocity $c/3$, slows and becomes even denser
 during the cooling process. In the extreme relativistic limit the pressure 
 does not change during this process.

Combining the radiative flow solution with the strong shock jump
conditions we obtain:
\begin{equation}
\label{r_e_factor}
\tilde e_2 = 2 \Gamma^2 \rho_1,  
\end{equation}
\begin{equation}
\label{r_n_factor}
\tilde \rho_2 = 2 \sqrt{2} \rho_1 \Gamma {\sqrt{1+\epsilon} \over 1 - \epsilon},
\end{equation}
\begin{equation}
\label{r_gamma_factor}
\tilde \gamma_2^2 = { 1 \over 2}  \Gamma^2 \gf2.
\end{equation}


Using  Eqs. \ref{r_e_factor}, \ref{r_n_factor}
and \ref{r_gamma_factor}  we find  that $\epsilon$ is equal to the fraction
of energy  that each
particle has lost in the unshocked fluid frame 
during the cooling process:
\begin{equation}
\epsilon = { { 4 \gamma_2 e_2 /3 \rho_2 } - {4 \tilde \gamma_2 \tilde e_2 /3
    \tilde \rho_2}
   \over { 4 \gamma_2 e_2 /3 \rho_2  }}.
\end{equation} 
This definition of $\epsilon$ coincides with the definition of \cite{R1}.

The energy loss rate of the whole blast wave in the
unshocked matter rest frame is the difference between the work done on
the cooling layer by the internal pressure $\tilde p_2$, and the
increase in internal and kinetic energy of shocked matter. For 
spherical radiative blast wave we obtain:
\begin{equation}
\label{r_dE_dt}
{dE \over dt} = -4 \pi R_{sh}^2 \left[ \tilde p_2 \tilde \beta_2  - \left \{ (\tilde e_2 + \tilde p_2) \tilde \gamma_2^2
\tilde \beta_2  - \tilde p_2 \right \} \left \{
\beta(\Gamma) - \tilde \beta_2  \right \} \right] = -8 \pi R_{sh}^2
\Gamma^2 \epsilon /3.
\end{equation}
This rate equals the rate in which energy is supplied to the unshocked
matter ($4 \pi R^2 p_2$), multiplied by $\epsilon$.  \cite{bm}
have obtained an equation for ${dE / dt}$ in the limit $\epsilon
\to 1$ (Eq. 84 there) which differs from our result due to a missing
factor of $4/3$ 
in their Lorentz transformation of the energy density.
\footnote{ Energy density transforms like $e'=(e+p)\gamma^2-p$, which becomes $e'=(4/3)e\gamma^2$ in
the extreme relativistic limit. In this calculation \cite{bm} have used $e'=e\gamma^2$.
\label{four_thirds}}

It is important to note that there are many possible definitions of radiative efficiencies, which
are frame dependent. 
For instance, the fraction of internal energy lost in the shock frame is:
\begin{equation}
{ e_2 ( 4 {\gamma^s_2}^2-1)/ 3 \rho_2 \gamma^s_2 - \tilde e_2 ( 4 \tilde
  {\gamma^s_2}^2-1)/ 3 \tilde \rho_2 \tilde \gamma^s_2 \over 
 e_2 ( 4 
  {\gamma^s_2}^2-1)/ 3  \rho_2  \gamma^s_2} = 
{ \epsilon(37+\epsilon (16+3\epsilon)) \over 7 (1+\epsilon) (3+\epsilon)}.
\end{equation}
This equals $\epsilon$  for   $\epsilon=0$ and  $\epsilon=1$, but it has a
different form for intermediate values.

\begin{figure}
\begin{center}
\plotone{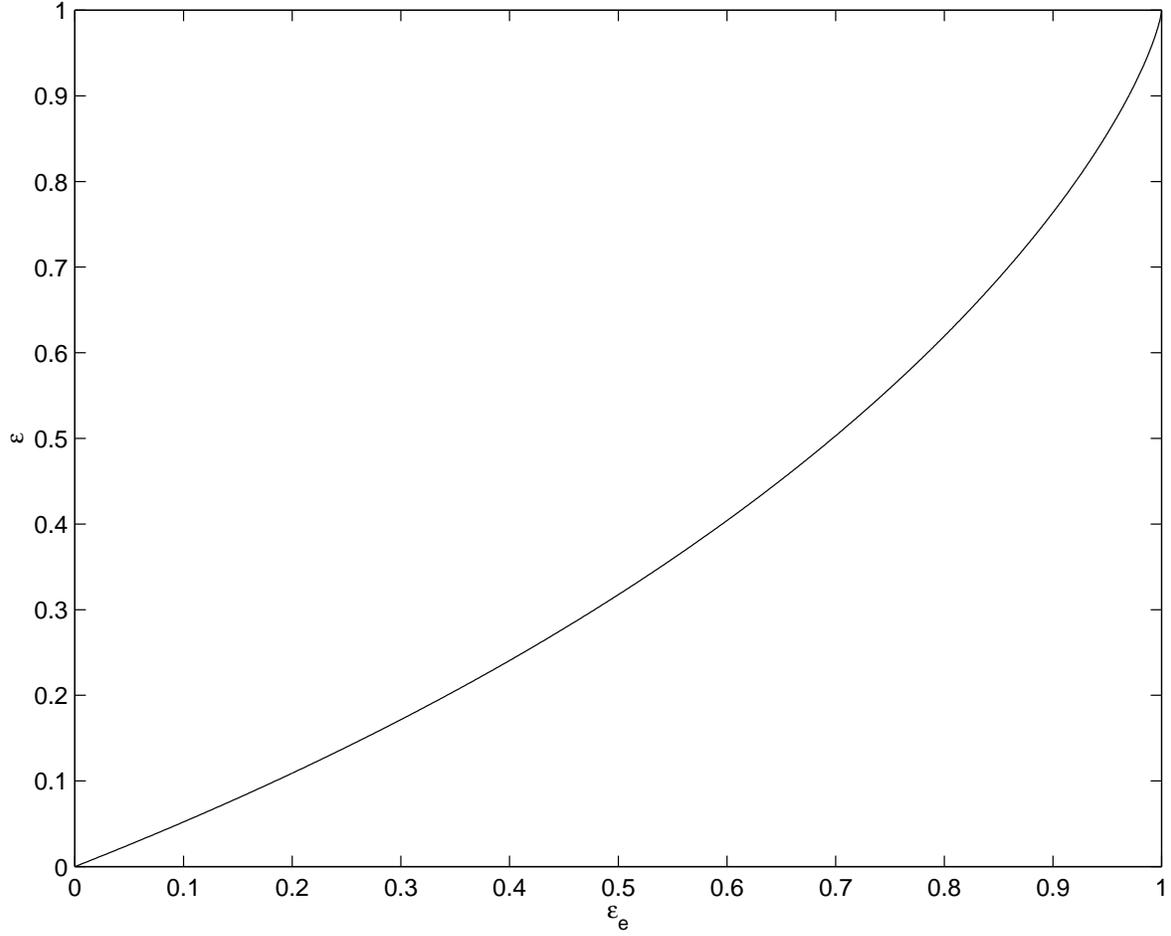}
\caption{ 
 Radiated energy as a fraction of the work done by the shock on the
 surrounding medium, vs. the fraction of internal energy
 the shock distributes to electrons. The calculation assume that electrons and
 protons are not coupled outside of the shock, and that only the electrons radiate.
\label{fig:eps_e}
}
\end{center}
\end{figure}

A physical example of a partially radiative shock wave leads to  another
definition of radiative efficiency. 
Consider a system in which the shock distribute the energy
between electrons and protons according to:
\begin{equation}
e_2^{elec}=\epsilon_e e_2, \quad e_2^{prot}=(1-\epsilon_e) e_2,
\end{equation}
and  there is no coupling between electrons and protons afterwards. 
Charge neutrality requires that the density of electrons and protons remains equal. 
The protons do not radiate.
Therefore their flow is adiabatic. When the electrons have cooled
down completely,
the protons' energy satisfy:
\begin{equation}
\tilde e_2^{prot} \tilde \rho_2^{-\pgamma} = e_2^{prot}  \rho_2^{-\pgamma}.
\end{equation}
At this stage the cold electrons have no internal energy, $\tilde
e_2^{elec}=0$ and  therefore  $\tilde e_2=\tilde e_2^{prot}$.
Using Eq. \ref{r_n_factor} we obtain:
\begin{equation}
\epsilon_e = 1 - {\left({ 1- \epsilon \over \sqrt{1+\epsilon}} \right )}^{\pgamma}.
\end{equation}
Again, $\epsilon_e$ equals $\epsilon$ for $\epsilon=0$ and for $\epsilon=1$ (see Fig.
\ref{fig:eps_e}), but it has a different form. It is simple to understand
this by recalling the behavior in the cooling layer. As a fluid element
cools it {\it slows} in the shock frame and {\it accelerates} in the
observer frame. The electrons 
accelerate the protons while cooling in order to maintain charge neutrality. This
acceleration transfers energy from the electrons to the protons.
Therefore, not all the electrons internal energy is radiated away.

\subsection{Self similar solution for homogeneous medium}
\label{s:homo}
Inside the blast wave the flow is adiabatic and
the equations of  motion
are obtained
by setting the divergence of the  energy momentum tensor
to zero (see e.g. \cite{ll} pg. 506 ). For spherically symmetric flow, the
pressure, energy and velocity satisfy
\begin{equation}
\label{r_spheri_particle}
{\partial \gamma \rho \over \partial t}+{1\over r^2}
{\partial \over \partial r} r^2 \rho \gamma \beta =0
\end{equation}
\begin{equation}
\label{r_spheri_momentum}
{\partial \over \partial t} \gamma^2 ( e+\rho+p) \beta+ {1\over r^2}
{\partial \over \partial r} r^2 \gamma^2 ( e+\rho+p) \beta^2 =0
\end{equation}
\begin{equation}
\label{r_spheri_energy}
{\partial \over \partial t} \gamma^2 ( e+ \rho + \beta^2 p)+ {1\over r^2}
{\partial \over \partial r} r^2 \gamma^2 ( e+\rho+p) \beta =0
\end{equation}
where $(r,t)$ are in the unshocked matter frame and $r=0$ is
the center of the blast wave.

As in the Newtonian case we look for a solution in which the
energy
decreases as a power-law in time:
\begin{equation}
\label{G_of_t}
\Gamma^2 \propto t^{-m}, \quad m > 3.
\end{equation}
Following \cite{bm}, we define a similarity variable
\begin{equation}
  \label{r_chi_def}
  \chi = [ 1+2(m+1)\Gamma^2(t)](1-r/t).
\end{equation}

We substitute the self similar variables and take the  extreme relativistic
limit
\footnote{ 
While solving Eqs. \ref{r_spheri_particle} and \ref{r_spheri_energy}
in the extreme relativistic limit 
we assume  $\rho \ll e$.
We check the validity of this limit
by requiring that the flow continues to be extreme relativistic after
the cooling layer,
${\tilde e_2 / \tilde \rho_2 \gg 1}$.
This results in a requirement $\Gamma \gg {1/(1- \epsilon)}$. It
means that as the blast wave looses more energy, our solution breaks down, even though  $\Gamma \gg 1$.
}
of 
Eqs. \ref{r_spheri_particle}, \ref{r_spheri_momentum} and       \ref{r_spheri_energy}.
We have to expand Eqs.  \ref{r_spheri_momentum} and
\ref{r_spheri_energy} to the second order in $\Gamma^{-2}$ as these 
equations are identical in the leading order in $\Gamma^{-2}$.

We  write the pressure, velocity, and
density of the shocked fluid as
\begin{equation}
\label{r_f_def}
p(r,t) = {1 \over 3} \tilde e_2 f(\chi) = {2 \over 3} \rho_1 \Gamma(t)^2 f(\chi),
\end{equation}
\begin{equation}
\label{r_g_def}
\gamma(r,t)^2 = \tilde \gamma_2^2 g(\chi) = {1 \over 2}  \Gamma(t)^2 \gf2 g(\chi),
\end{equation}
\begin{equation}
\label{r_h_def}
\rho(r,t) \gamma(r,t) = \tilde \gamma_2 \tilde \rho_2  h(\chi ) = 2 \rho_1 \Gamma(t)^2 { 1+\epsilon \over 1- \epsilon} h(\chi).
\end{equation}
These expressions are valid for the interior of the blast wave, $\chi \geq
1$. The boundary conditions at $\chi=1$ (Eqs. \ref{r_e_factor},
\ref{r_n_factor} and \ref{r_gamma_factor}) are satisfied by setting
\begin{equation}
f(1)=g(1)=h(1)=1.
\end{equation}

Substituting  Eqs. \ref{r_f_def}, \ref{r_g_def} and \ref{r_h_def} 
in       Eqs. \ref{r_spheri_particle},
\ref{r_spheri_momentum} and 
\ref{r_spheri_energy},    
and replacing the
derivatives using
\begin{equation}
{d\Gamma(t) \over dt} = -{m \over 2} { \Gamma(t) \over t },
\end{equation}
\begin{equation}
t {\partial \chi \over \partial t} = (m+1)(2 \Gamma^2 - \chi) + 1,
\end{equation}
and
\begin{equation}
t {\partial \chi \over \partial r} = -[ 1 + 2(m+1)\Gamma^2]
\end{equation}
all to the second  order in $(\Gamma^{-2})$,
we obtain  the following  self similar  equations:
\begin{eqnarray}
\label{r_ss_g}
{A \over g} {d \ln g \over d \chi} &=&
   { (7m-4)-(m+2) \gf2 g \chi  } \\
\label{r_ss_f}
{A \over g} {d \ln f \over d \chi} &=&
  { 8(m-1)-(m-4) \gf2 g \chi  } \\
\label{r_ss_h}
{A \over g} {d \ln h \over d \chi} &=&
  \Big( 2(9m-8)-2(5m-6) \gf2 g \chi + \\ & & (m-2)\gf2^2 g^2 \chi^2 \Big)/ 
(2-\gf2 g \chi), \nonumber
\end{eqnarray} 
where
\begin{equation}
\label{def_A}
 A={ (m+1)(4-8 \gf2 g \chi + \gf2^2 g^2 \chi^2) \over \gf2}.
\end{equation}

To solve these equations we need a relation between the amount of
cooling, 
presented in the equations by the parameter $\epsilon$ and the
evolution parameter $m$.
The energy stored in the blast wave is given by:
\begin{equation}
\label{r_energy}
E(t) = \int_0^{R(t)} { 4 \pi \over 3}  r^2  e (4\gamma^2(r,t)-1)  dr = 
{ 8 \pi \rho_1 \over 3(m+1) } \Gamma^2 t^3 \gf2 \int_1^{\infty} f g d\chi,
\end{equation}
to the leading order in $\Gamma^{-2}$.
The time derivative of the energy satisfies
\begin{equation}
\label{r_d_energy}
{dE \over dt} = (3-m) {E \over t}.
\end{equation}

Combining Eqs. \ref{r_dE_dt}, \ref{r_energy} and \ref{r_d_energy}  we
obtain a normalization equation which  combines the hydrodynamic profile
through $f$ and $g$, the evolution parameter $m$, and the 
energy loss parameter:
\begin{equation}
\label{r_energy_cond}
1 - {m-3 \over m+1} \int_1^{\infty} f g d\chi = {1 \over 1 + \epsilon}.
\end{equation}

\begin{figure}
\begin{center}
\plotone{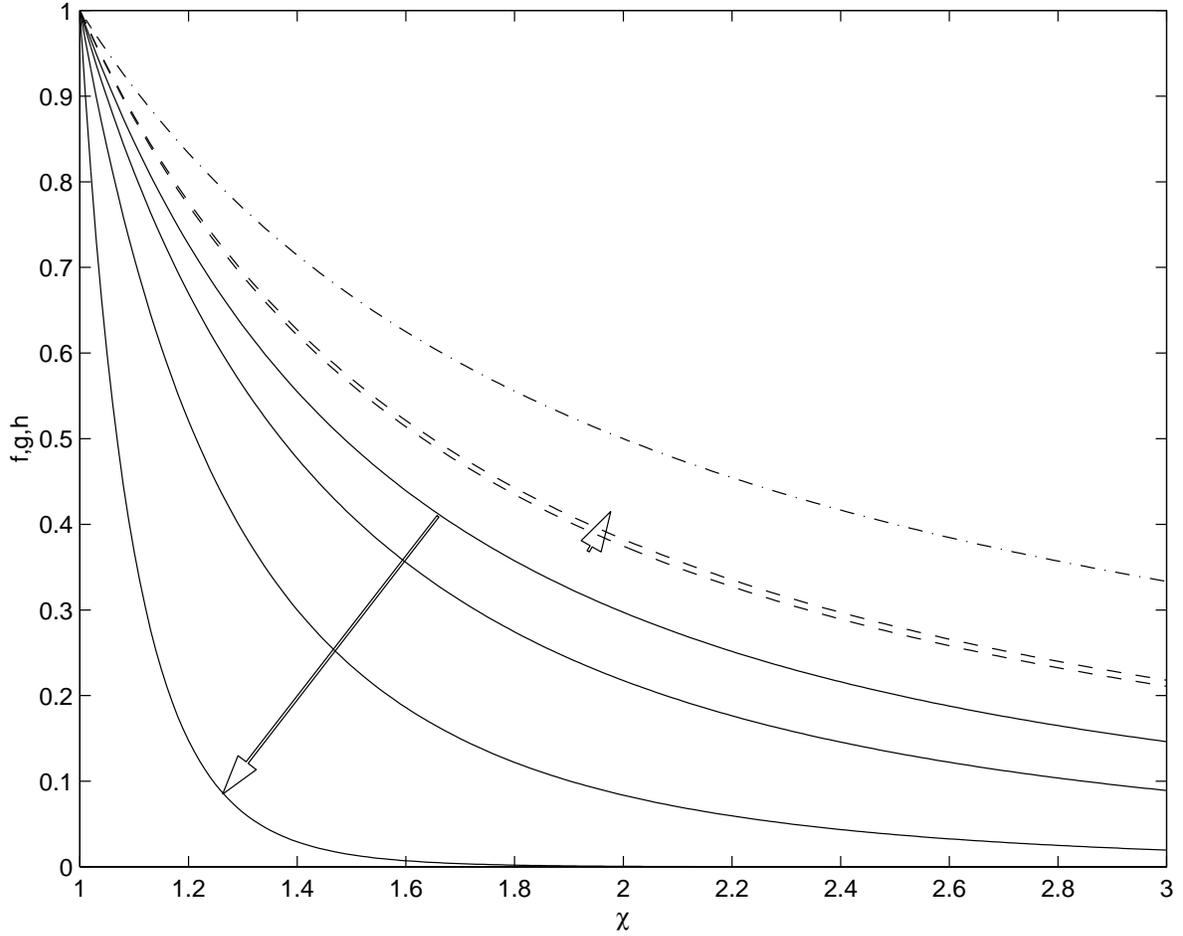}
\caption{ 
The normalized density (solid), the Lorenz factor (dashed-dotted)  and
the pressure
(dashed)  profiles of the self similar extreme relativistic solution, for
$\epsilon=0,0.5,0.8,0.95$. The arrows shows the direction of increasing $\epsilon$.
\label{rel_profile}
}
\end{center}
\end{figure}

Surprisingly, there is an analytic solution to Eqs. \ref{r_ss_g}, \ref{r_ss_f},
\ref{r_ss_h} and \ref{r_energy_cond} for arbitrary values of  $\epsilon$: 
\begin{equation}
\label{r:m_of_eps}
m = { \epsilon^2+14 \epsilon + 9 \over 3 - \epsilon} 
\end{equation}
\begin{equation}
g = \chi^{-1}, \quad f=\chi^{ -\alpha_1}, \quad h=\chi^{-\alpha_2},
\end{equation}
\begin{equation}
\alpha_1=1 + { 5 \over 12+\epsilon}, \quad \alpha_2={\epsilon^2+14 \epsilon
  -21 \over \epsilon^2+11 \epsilon -12 }
\end{equation}
and 
\begin{equation}
E(t) = { 8 \pi \rho_1 \over 3 }  { 3 - \epsilon \over 17+\epsilon} \Gamma^2 t^3.
\end{equation}
The functions $f,g,h$ are displayed in Fig. \ref{rel_profile}.
In the limit of an adiabatic blast wave,
$\epsilon \to 0$, we recover  the Blandford-McKee
solutions. The matter is concentrated, in this solution, in  a narrow shell of width
$R/\Gamma^2$.  As  $\epsilon$ increases, the matter tends to
form even narrower, denser shell. Similarly to the Newtonian
solution, the internal energy does not decrease as fast as the matter
density in the interior of the blast wave. In fact,   
the width of the pressure profile is approximately independent of
$\epsilon$. 

Our solutions are expressed in the un-shocked matter frame, which is
the same as the observer frame.  However, due to the relativistic
motion of the emitting matter toward the observer, the time difference
between the arrival of  two photons to an observer at infinity is not equal
to the delay between their emission.
A photon released at
time $t$ from the radiation layer will reach the observer at a time 
\begin{equation}
t_{obs} = t - R(t) = { t \over 2 (m+1) \Gamma^2} \propto t^{m+1}.
\end{equation}

All the observables (luminosity, frequency, etc.) of blast waves should,
of course, be given in the observer time. 
For instance, the bolometric luminosity is the derivative of 
the blast wave energy to the observer
time. Using  Eq. \ref{r_d_energy} we find 
\begin{equation}
\label{m_to_obs}
L \propto {dE \over dt_{obs}}  \sim t_{obs}^{-{m-3 \over m+1}-1}.
\end{equation}

\begin{figure}
\begin{center}
\plotone{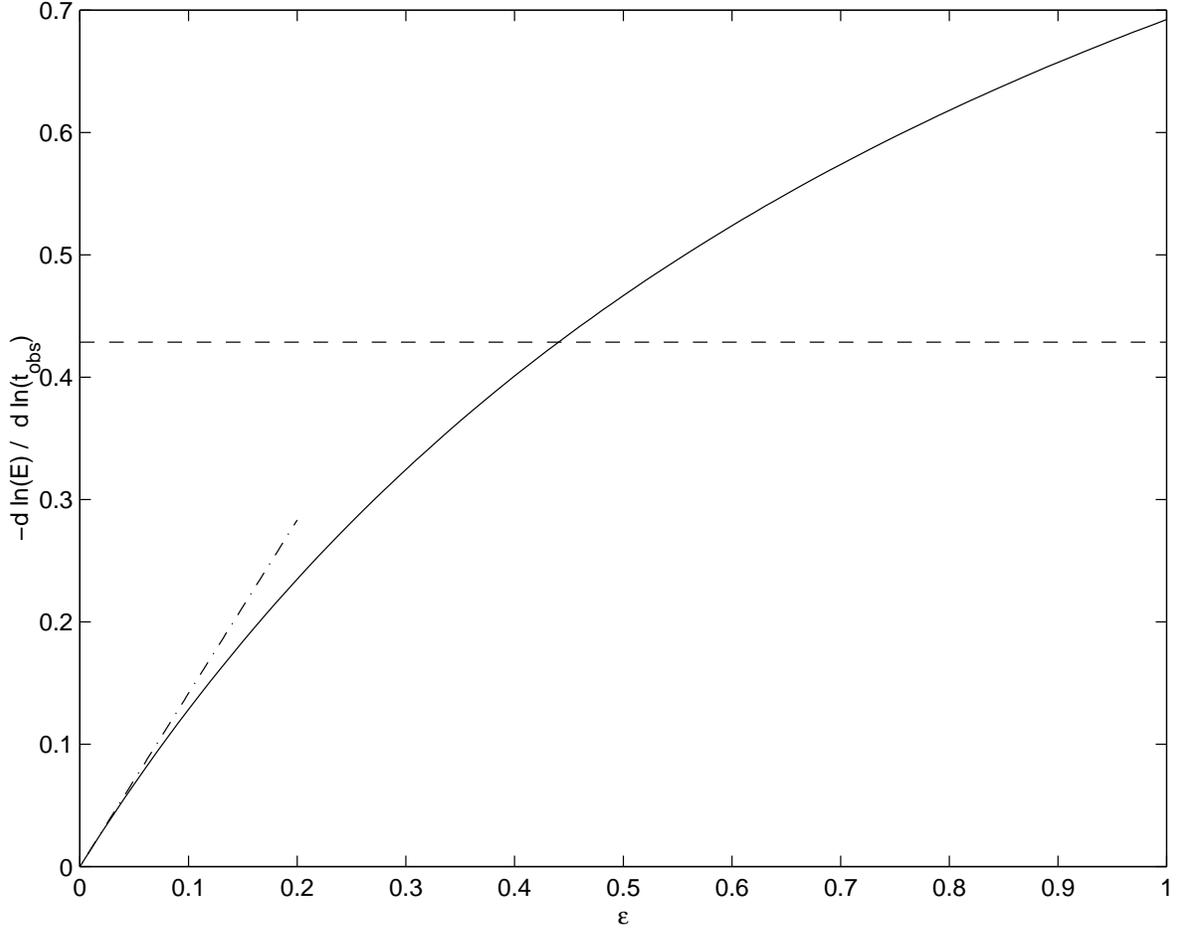}
\caption{ 
  The observed energy loss rate, $- d \log(E) / d \log(t_{obs})$, as a
  function of $\epsilon$, extreme relativistic case (solid line).  The
  dashed line is the energy loss rate for the fully radiative, MCS
  like, Blandford-McKee solution. The dashed-dotted lines are the
  linear approximation for $\epsilon \to 0$, assuming that the self
  similar profile does not depend on $\epsilon$
\label{loss_rel}
}
\end{center}
\end{figure}

An interesting case is a sudden release of all the blast wave energy.
As the energy is released, the 
shell Lorentz factor also drops to unity. Due to the dependence of the observed time
on $\Gamma$, this event, no matter how short it is, will be spread over a long time.
Even if $m \to \infty$, the observed luminosity will not drop faster than
$L \propto t_{obs}^{-2}$. (In our solutions this limit
can not be reached because
$m<12$ for all $\epsilon$).
In Fig. \ref{loss_rel} we show the relation between the energy drop rate
and $\epsilon$.

\begin{figure}
\begin{center}
\plotone{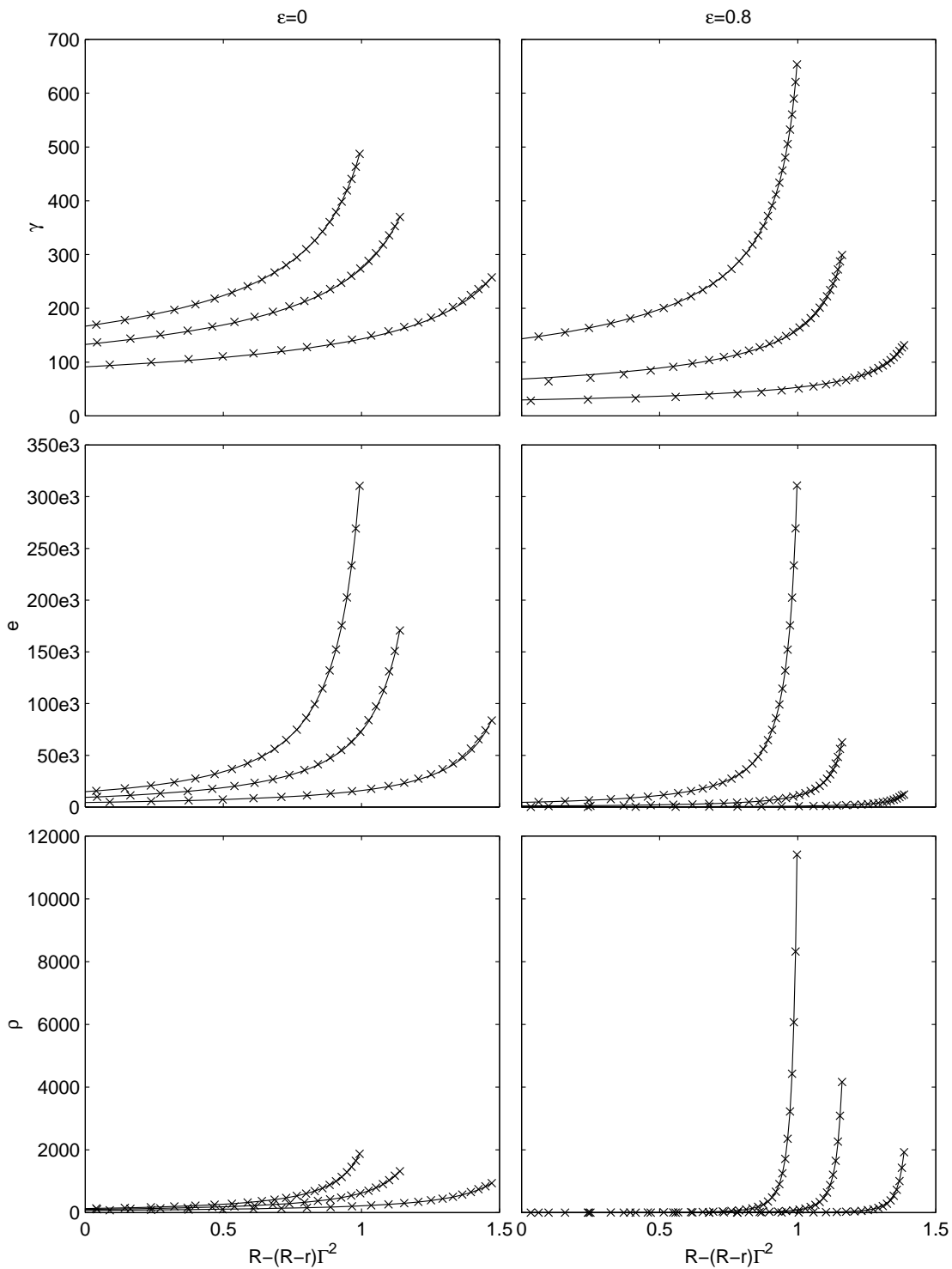}
\caption{ 
  Simulated (x-marks) and self similar (solid line) solutions for the
  {\it  interior } of an adiabatic ($\epsilon=0$) and a partially radiative
  ($\epsilon=0.8$) blast waves. The initial conditions are
  $\Gamma=\sqrt{2} \cdot 500$, $R=1$. Note the larger Lorentz factors and
  densities in the radiative solution.
\label{fig:simulation}
}
\end{center}
\end{figure}

We check our solution for the blast wave interior using a spherical
one dimensional numerical simulation. For simplicity, to avoid detailed modeling
of the cooling process, we use our analytic solution for the
``radiative shock''. At each time step we calculate the location we
expect the shock to be, assuming that the evolution follows our self
similar solution.  We set the values of density, energy and velocity
at that location using the modified shock conditions of Eqs.
\ref{r_e_factor}-\ref{r_gamma_factor}. As the  initial
conditions we use our self similar solution, and check if the
calculated profiles follow it. We continue until the solution 
becomes non
extreme relativistic. We find a good agreement between
the numerical profiles and the self similar ones, see Fig.
\ref{fig:simulation}.  

\cite{bm} have also found a self similar solution for 
blast waves with injection of energy.
This solution contains an  additional internal shock wave.
It is interesting to check whether it is possible to find another radiative solution 
by incorporating such a  shock wave into our self similar solution.
To check this, we write the velocity of a sphere with
constant $\chi$ by inverting Eq. \ref{r_chi_def} to $r=r(\chi,t)$ and
applying  a partial time  derivative. We obtain:
\begin{equation}
\Gamma_{\chi}^2=\Gamma^2 / \chi.
\end{equation}
The velocity difference between the fluid and this sphere  is:
\begin{equation}
\label{gamma_diff}
\Gamma_{diff}={3+\epsilon \over 2 \sqrt{2 \gf2}},
\end{equation}
which is constant over the whole profile.  We find that for all
possible values of $\epsilon$, the Lorentz factor of the velocity
difference $\Gamma_{diff}<\sqrt{3/2}$ (the local speed of sound), i.e.
the flow is subsonic, while a shock requires supersonic flow.
Therefore an additional shock wave cannot take place within this self
similar solution.

\subsection{External medium with a density gradient}
These solutions, described in  Sec. \ref{s:homo}, can be generalized to the
case where the external medium density has a power law profile.
If the density gradient is $\rho_1 \propto r^{-k}$, we have
\begin{equation}
\rho_1 \propto \Gamma^{2k/m}, \quad E \propto \Gamma^{2+2k/m}R^3 \propto R^{-(m-(3-k))}.
\end{equation}
The  self similar are:
\begin{eqnarray}
\label{density_grad_d}
{A \over g} {d \ln g \over d \chi} &=&
   { (7m+3k-4)-(m+2) \gf2 g \chi  } \\
{A \over g} {d \ln f \over d \chi} &=&
   { 8(m-1)+4k-(m+k-4) \gf2 g \chi  } \\
{A \over g} {d \ln h \over d \chi} &=&
    \Big( 2(9m+5k-8)-2(5m+4k-6) \gf2 g \chi + \\ & & (m+k-2)\gf2^2 g^2 \chi^2 \Big)/ 
(2-\gf2 g \chi), 
\label{density_grad_h}
\end{eqnarray} 
where A is defined by Eq. \ref{def_A}.
Equations \ref{density_grad_d}-\ref{density_grad_h}
have an  analytic solution:
\begin{equation}
g = \chi^{-1}, \quad 
f=\chi^{-\alpha_1}, \quad
h=\chi^{-\alpha_2} 
\end{equation}
\begin{equation}
\alpha_1 =  { 17 + \epsilon - 4 k \over 12 + \epsilon - 3 k},\quad
\alpha_2 = { 21 - \epsilon ( 14+\epsilon-4k)-6k   \over (1-\epsilon) (
  12 + \epsilon -3k)}
\end{equation}
for 
\begin{equation}
m = {\gf2^2+  3 \gf2 (4-k) -4  \over 3-\epsilon} > -1.
\end{equation}
The energy contained in the blast wave is therefore given by:
\begin{equation}
E(t) = { 8 \pi \rho_1 \over 3 }  { 3 - \epsilon \over
  17+\epsilon-4 k} \Gamma^2 t^3,
\end{equation}
where $\rho_1$ should be understood as the density of the external
medium at the position of the shock.

As in the solution with homogeneous medium, it is interesting to check if another
solution can be obtained by incorporating another shock into the self similar solution.
However,  Eq. \ref{gamma_diff} is valid also for the this solution, 
and no additional shock can be fitted into this self-similar solution.

\section{Comparison to the limiting radiative cases}
\label{s:others_results}
\subsection{The fully radiative solution}
\label{s:eps_1}

The fully radiative solution deserves a special attention due to the
singularity in the boundary conditions both in the Newtonian and in
the extreme relativistic limits.  To clarify the situation we
re-derive following \cite{OM88} the Newtonian fully radiative
solutions.  Then we discuss the validity of these solutions and
compare them to our self similar solution.  Next we obtain a new fully
radiative solution for the relativistic case and compare it to the
Blandford-McKee fully radiative solution and to our self similar
solutions.

In the Newtonian case we assume that
all the matter is concentrated in a narrow shell, and look for a self
similar solution. We define a dimensionless kinetic energy, and use
Eq. \ref{n_beta_xi} with $\epsilon=1$ to find
\begin{equation}
\label{non_dim_kinetic}
  \sigma = { \rho_1 V U_{sh}^2 \over E_0 ({t \over t_0})^{\lambda} } = - {2 \lambda \over 3 }({5 \over 2+\lambda}),
\end{equation}
where $V= 4 \pi R_{sh}^3 / 3$ is the volume swept  by the blast wave.
The pressure must be constant in the empty interior (otherwise it will
lead to infinite acceleration). We find 
\begin{equation}
\label{non_dim_P}
  \bar P = { (\pgamma -1) \over V} ( E_0 ({t \over t_0})^{\lambda}  - { 1 \over 2}
  \rho_1 V  U_{sh}^2) =  \rho_1 U_{sh}^2 ( {1 \over \sigma} - {1 \over 2}) (
  \pgamma -1 ). 
\end{equation}
The equation of motion for a narrow shell is:
\begin{equation}
\label{non_dim_motion}
  {d (\rho_1 V U_{sh}) \over dt} = 4 \pi R_{sh}^2 \bar P.
\end{equation}
Substitution of  Eqs. \ref{non_dim_kinetic} and
\ref{non_dim_P} in Eq. \ref{non_dim_motion} yields a quadratic equation with two solutions:
\begin{equation}
  \lambda = \left\{ \matrix{ -{3\over 4} & MCS \cr 
-{ 6 (\pgamma -1 ) \over 2 + 3 \pgamma} & PDS } \right. 
\end{equation}
The MCS solution incorporate an empty interior, with no pressure.
In the PDS solution
\begin{equation}
\bar P V^{\pgamma} \propto { R_{sh}^2 \over t^2} R_{sh}^{3 \pgamma}
\propto t^{ \left( {2+\lambda \over 5 } \right) (2+3 \pgamma)-2} = const.,
\end{equation}
during the blast wave expansion corresponding to an adiabatic expansion
in which new matter does not enter the blast wave interior. This
behavior is consistent with our assumption that new matter accretes
only on the expanding shell.

The existence of two possible solutions with the same boundary conditions 
is unusual. Therefore, one should recognize 
the physical conditions which leads to each of the self similar solutions.
\cite{CMB88} have found numerically that a supernova remnant (SNR)
evolves initially  according to the adiabatic Sedov-Taylor solution,
becomes PDS when the energy loss near the shock becomes important, and
finally becomes MCS when other processes cool the blast wave interior.
\cite{G83} has investigated blast waves with power-law cooling functions,
${\cal L} \propto \rho T^{-c}$. He has found that $c>-2/3$ leads to the MCS solution,
$c \to -2$ leads to the PDS solution and other values of $c$ do not lead 
to self similar solutions.
Since lower values of $c$ correspond to less effective cooling in lower temperature
(i.e., to a more adiabatic interior), no interior cooling results, asymptotically, in the 
PDS solution.

Approaching the limit $\epsilon \to 1$, without taking into account
interior cooling, results in the PDS solution. During the PDS
expansion no matter enters the interior which expands adiabatically.
Incorporating even an infinitesimal interior cooling will cool the
interior down eventually, and the blast wave will asymptotically
approach the MCS solution.  Therefore, the limit of a fully radiative
blast wave with adiabatic interior results in the MCS or PDS solution
depending on the way we reach the limit of an adiabatic interior..
Our self similar solution reaches the PDS solution in the $\epsilon
\to 1$ limit.  This is due to the adiabatic  interior  assumption,
which, following the previous arguments, forbids a MCS solution.
 
\cite{bm} have calculated the dynamics of a fully radiative relativistic blast wave
with an impulsive supply of energy, treating the swept-up matter as
lying in a thin, cold shell adjacent to the shock and assuming a cold
interior (similarly to the Newtonian MCS solution). 
They have found that $\Gamma \propto R^{-3}$, which translates to an
evolution of energy as seen by the observer:
\begin{equation}
E \propto t_{obs}^{-3/7}. 
\end{equation}

We proceed by deriving a PDS like relativistic solution.
In the extreme relativistic limit we assume that $e \gg \rho$ in order
to obtain self-similar solution. The motion of a matter shell is
not self-similar, and we cannot follow the Newtonian derivation step by step.
However, we know from Eq. \ref{gamma_diff} that in the $\epsilon \to 1$ limit the matter
moves with same speed as shell of constant $\chi$, i.e. new matter does not enter the 
blast wave interior, as in the Newtonian solution. 
We obtain the internal pressure, $P$, looking at the evolution of 
a shell with a constant $\chi$ and a constant width $d \chi$.
Using Eq. \ref{r_chi_def} we find to the leading order in $\Gamma^{-2}$:
\begin{equation}
dR \propto d \chi \Gamma^{-2-2/m}, \quad R \propto \Gamma^{-2/m},
\end{equation}
where $dR$ and $R$ are the shells' width and its radius, in the unshocked matter frame.
We use Eq. \ref{r_f_def} to find that $P \propto \Gamma^2$, and obtain:
\begin{equation}
P V^{\pgamma} \propto \Gamma^{2-(6/m+1)\pgamma},
\end{equation}
including a Lorentz transformation of the shell width to the matter frame.
Requiring that $P V^{\pgamma}=const.$ we obtain $m=12$. 
Using Eq. \ref{m_to_obs} we finally find:
\begin{equation}
E \propto t_{obs}^{-9/13},
\end{equation}

As in the Newtonian case, our self similar relativistic solution approaches the PDS solution
in the fully radiative limit. An analytic treatment of the validity of the MCS and PDS
solution does not exist in the relativistic case. However, in view of the Newtonian solution
we expect that interior cooling is needed in order to obtain the MCS solution. 

\subsection{Solutions for  $\epsilon \ll 1$}
Using the conditions at the shock, and our  radiation model
with parameter $\epsilon$, we have found the energy loss rate for
the Newtonian (Eq. \ref{n_dE_dt}) and for the extreme relativistic
(Eq. \ref{r_dE_dt}) limits.  Assuming that a self similar
solution exists, we have found scaling laws for the conditions at the
shock, (Eq. \ref{n_shell} or Eq. \ref{G_of_t}).
Substituting the  energy (calculated from the self similar
profiles, Eqs \ref{n_E_of_t} or \ref{r_energy} ) and combining the two, we have found
equations which relates the energy decrease rate, the radiative
parameter and the hydrodynamic profiles (Eqs.
\ref{n_beta_xi} or Eq. \ref{r_energy_cond}).

Self similarity requires  that the energy is a power-law of time.
The non-dimensional profiles are needed to
determine the power-law index.  
In Sec. \ref{s:newtonian_self} and Sec. \ref{s:homo} of
this paper, we have  solved  the self similar equations to
obtain this index. However, as suggested by 
the relativistic calculations of \cite{R1}, if $\epsilon \ll
1$ a simpler derivation exists. We assume that in this case the
radiation does not alter the hydrodynamic profiles and use the
adiabatic   Sedov-Taylor or Blandford-McKee profiles to obtain the
power-law index.

In the Newtonian case the hydrodynamic profiles enter
Eq. \ref{n_beta_xi} through the self similar location of the shock
front. Assuming that the profiles do not depend on radiation, the
shocks' location is constant, i.e.  $\xi_0=\xi_0({\epsilon=0})$.
Substituting this into 
Eq. \ref{n_beta_xi}  we  find the energy decrease rate to the lowest order
\begin{equation}
  \label{n_beta_xi_inf}
  \lambda=-{ 16 \pi {[\xi_0({\epsilon=0})]}^5 \over 125} \epsilon + O(
  \epsilon^2).
\end{equation}
For $\pgamma=5/3$ we obtain $\lambda=-0.81 \epsilon + O( \epsilon^2
)$. The limiting curves appear in Fig. \ref{loss_newt}.

In the relativistic case the profiles enters Eq. \ref{r_energy_cond}
through $\int_1^{\infty} f g d\chi$. We use the adiabatic Blandford-McKee
solution to obtain $\int_1^{\infty} f g d\chi=12/17$, and find
to the
lowest order in $\epsilon$
\begin{equation}
  \label{r_beta_xi_inf}
 - {d \log(E) \over d \log(t_{obs})} ={1 \over \int_1^{\infty} f g d \chi} \epsilon +
  O(\epsilon^2) = {17 \over 12} \epsilon + O( \epsilon^2).
\end{equation}
This limit is shown in Fig. \ref{loss_rel}. 
This result differs from 
\cite{R1} due to a missing factor of $4/3$ in his Lorentz
transformation of the energy density (see footnote \ref{four_thirds}).

We conclude by noting that in the Newtonian and in the extreme
relativistic limits, the change in the blast wave evolution causes the
energy to decrease slower  than what was calculated earlier using
adiabatic estimates.

\section{Conclusions}
We have solved the hydrodynamical evolution of 
blast waves in which each shocked particle emits a fixed
fraction of the energy it gains in the shock.
We have divided  the blast wave into three regions: the adiabatic
shock, the radiative layer, and the adiabatic interior. 
The adiabatic shock and the radiation layer were  combined to form a 
``radiative shock'', which set the boundary conditions for the adiabatic interior.

For fast cooling cases we have obtained a solution for a planar radiative
layer with arbitrary shock velocities, independent of the cooling
process. We have found that in the shock frame the fluid cools, 
slows down and its pressure {\it increases} during the cooling process.

We have obtained  self similar solutions for adiabatic interior 
for the Newtonian and
the extreme relativistic cases. We find that radiation can change the
hydrodynamics considerably. Because of this change in the blast wave evolution
the luminosity  decays  slower (in time) than what was estimated earlier
assuming that the self similar profiles are independent of the
radiated energy.

We have also found that as a blast wave becomes more radiative, its
matter concentrates near the shock and forms a dense shell. However,
the internal pressure does not drop to zero. In the fully radiative
limit of Newtonian blast waves we have reached the pressure-driven
solution. We have obtained a new extreme relativistic solution for
fully radiative blast wave, which resembles the Newtonian PDS
solution.  Our self similar solutions reach this modified solution in
the fully radiative limit.  This solution does not correspond to the
Blandford-McKee radiative solution.

The pressure must be  continuous within the adiabatic interior and it
cannot develop self a similar shock or
a rarefraction wave (apart from the main strong shock with the ISM). 
Therefore, even in the Newtonian PDS limit with isobaric 
interior, the solution contains a self similar transition
layer between the ``radiative shock'' conditions and the internal
pressure.

The recently discovered X-ray, optical and radio emission following a
GRB, so called ``Afterglow'' is widely believed to be the result of
the deceleration of a relativistic material that collides with the
surrounding matter.  According to the common model, the shock wave
produced by the collision accelerates electrons to relativistic
velocities. These electrons, that carry a fixed fraction of the
internal energy produced by the shock emit synchrotron radiation which
is the observed afterglow. For reasonable parameters, the electron
cooling is fast (at least in the early stages of the afterglow
evolution), so that the electrons loose most of their energy. Since
the density behind the shock is low, the coupling between the
electrons and the protons is negligible.  Therefore, the protons
energy is not radiated away. This leads to a partially radiative
rather than a fully radiative shock, followed by an adiabatic flow.
This is {\it exactly} the scenario that leads to the partially
radiative blast waves derived in this paper. Our new self similar
partially radiative blast wave can serve as the basic hydrodynamic
solution from which spectra and light curves of the afterglow can be
calculated.  In particular we derive a new   $\Gamma(R)$ relation 
(Eqs. \ref{r:m_of_eps},
\ref{G_of_t} ) between the Lorentz factor and
the radius. This is the critical relation that determines most
afterglow observations.

This research was supported by a US-Israel BSF grant 95-238  
and by a NASA grant NAG5-3516. Re'em Sari thanks the Clore Foundations
for support.

\eject

\end{document}